**Dendritic Oxide Growth in Zero-Valent Iron Nanofilms Revealed by Atom Probe Tomography**

Mavis D. Boamah,[1] Dieter Isheim,[2] and Franz M. Geiger[1,*]

[1]Department of Chemistry, Northwestern University, Evanston, IL 60208, [2]Department of Material Science, and Center for Atom Probe Tomography, Northwestern University, Evanston, IL 60208, USA

**Abstract.** Atom probe tomography (APT) analysis of chemically pure nanofilms of zero-valent iron (Fe(0), or ZVI) and their thermal oxide nano-overlayers reveals the presence of dendritic iron oxide features that extend from the oxide nano-overlayer surface into the ZVI bulk. The dendrites are observed by APT to be in the 5 nm x 10 nm size range and form quickly under natural atmospheric conditions. Their growth into the ZVI lalyer is, within the limit of our three-month long study, self-limiting (i.e. their initial growth appears to quickly discontinue). The atomistic views presented here shed first light on the atmospheric corrosion process of Fe(0)-bearing engineered nanostructures and their surfaces in the limit of low bulk impurities. Possible roles of the newly identified oxidized iron dendrites are also discussed in the context of passivation processes limiting technological applications of Fe(0).

*Corresponding author: geigerf@chem.northwestern.edu

**I. Introduction.** Films of Fe(0) having nanometer thickness have potential applications in environmental remediation,[1] catalysis,[2] optics,[3] data storage,[4] consumer electronics and coatings products, and can serve as models for corrosion studies[5] of Fe(0)-bearing engineered building and construction materials. Of particular importance for the continued and sustainable supply of clean and safe water, the formation of passivation layers of oxidized iron over zero-valent iron (ZVI) materials can significantly curtail the long-term effectiveness of the permeable reactive barriers (PRBs)[6-12] that are widely used to remediate legacy pollution of ground water by toxic metals. Zero-valent iron is known to oxidize quickly under ambient conditions to Fe(II) and Fe(III) species.[1,13] The passivation layer has a defective nanocrystalline microstructure that can feature well-defined grain boundaries in aqueous environments.[14] Passivation of metals can also occur through penetrative oxidation resulting from crevices, cracks, or stress corrosion.[15-16]

Unfortunately, the rejuvenation of corroded iron-bearing engineered materials, or the ZVI particles used in PRBs, are curtailed, in part, by the lack of microscopic information (spatially resolved chemical information) on what chemical species passivate the active iron-bearing materials. Scanning tunneling microscopy and various X-ray techniques have been used to characterize chromium adsorption/reduction at iron oxide interfaces and the structure of passivated mineral oxides, but only with limited spatial resolution.[7,9,17-19] In contrast, atomic-scale studies such as atom probe tomography (APT) have the potential to reveal intrinsic properties of ZVI materials, in addition to structural features, that might influence their passivation. Most recently, APT has been used to map the isotopic distribution of iron when aqueous Fe(II) interacts with hematite.[20] Gault et al. probed a pure iron specimen exposed to $10^{-6}$ mbar of $O_2$.[21] Other recent APT studies have reported on nanostructures formed from penetrative oxidation in a nickel-chromium alloy.[16] Similarly, APT has been used to characterize the role of oxide penetration for the microstructure and propagation of crack tips in nickel-based alloy X-750 used in boiling water nuclear reactor stress-corrosion environments.[15]

Here, we employ APT to study zero valent iron nanofilms exposed to ambient air and to gain microscopic insights that we hope will be useful for elucidating pathways towards the rejuvenation of permeable reactive barriers. Moreover, the results presented here may help in the elucidation of the processes that lead to the corrosion of iron-based natural and engineered materials, and, more generally, of processes occurring at metal:oxide:water interfaces.

We previously reported a method for preparing nanometer-thin ZVI films using electron-beam physical vapor deposition (E-beam PVD) utilizing affordable iron sources (99.95% purity).[22-23] XPS analysis indicates an absence of common low-boiling point contaminants like sodium or zinc (*vide infra*). X-ray photoelectron spectroscopy (XPS) analysis indicates the presence of an ~5 nm thick oxidized iron overlayer, specifically Fe(III) on top of the bulk Fe(0) layer associated with a 25 nm ZVI film.[22] This overlayer forms spontaneously and rapidly when the ZVI specimen is exposed to air, and remains stable over prolonged (months and more) periods of time. While XPS, X-ray diffraction (XRD), and Raman spectroscopy provide a macroscopic understanding of this system, questions remain regarding the local structure, the local composition, and the spatial distribution of the iron oxide overlayer on top of the nm-thin ZVI films. Detailed structurally and chemically resolved information on the microstructure in the oxide nano-overlayer over ZVI nanofilms could serve as a conduit for overcoming ZVI passivation and optimizing the use of ZVI materials in heavy metal environmental remediation, construction, thin film applications, etc. Thus, the following questions arise:

1. How are Fe(III) and O atoms distributed in the ~5 nm thick oxide overlayer?

2. What is the nature of the boundary between the oxide overlayer and the Fe(0) bulk?

3. Why does Fe(III) appear in only the first ~5 nm layer and not below?

4. Does the iron oxide overlayer have any distinctive features that accelerate or hamper its growth into the bulk iron?

To answer these questions, we obtain and analyze three-dimensional atomic-scale views and detailed structural and elementally-resolved images of ZVI films reconstructed from APT. While scanning probes can provide excellent lateral resolution, these methods do not allow for resolving features below the surface, [24-25] and obtaining chemical information is challenging. Furthermore, spherical aberration-corrected scanning transmission electron microscopy generates a two-dimensional projection from which 3D information is difficult to obtain, and its sensitivity for chemical imaging is limited.[26-27] On the other hand, x-ray absorption near edge structure can provide information about the local bonding environments within the film, but falls short in giving us insight into the nano-interfacial structure, dislocations, and any nanofeatures that might be present.[28-29] As a nanoscale, element-specific 3D chemical imaging method, APT combines spatial resolution and chemical (element-, isotopic, and speciation-specific) information that allows us to obtain sub-nanometer detailed structural information and single-atom elemental analysis on nano-sized ZVI films in three dimensions. APT is advantageous specifically to probe subsurface and buried features such as precipitates, interfaces, and segregations in metals and alloys[21,30-35] with a spatial resolution of 0.3-0.5 nm. [36-38] In essence, as applied to our nanometer-thin ZVI films and their oxide nano-overlayers, APT enables the visualization and differentiation of the atomic spatial distribution within the iron oxide nano-overlayer and the ZVI nanofilm, and their various interfaces. While the method yields unprecedented views of the composition and structure of matter, both at the surface and the interior, we caution that APT has limitations in quantifying the exact amount of oxygen species present within oxides due to the isobaric peak overlap between the molecular ($O_2$) and atomic (O) oxygen species at a mass-to-charge-state ratio of 16 amu, and as a result of the potential formation of neutral oxygen species and atomic reconstruction in the course of sample field-evaporation that might not be detected.[39-40]

## II. Experimental

**A. Sample Preparation.** ZVI films are prepared via E-beam PVD deposition and have been previously described.[22] Briefly, we use iron boules of 99.95% purity as received from Kurt J. Lesker, Co. Prior to deposition of the ZVI film, silicon wafer substrates are cleaned by sonicating in methanol for 15 minutes and dried under ambient conditions. Detailed materials properties, including nanofilms hardness and composition by XPS, are provided in our prior work.[22] The ZVI films are exposed to ambient atmospheric conditions (air) for ~30 minutes before the surfaces are covered for specimen preparation and APT by depositing ~50 nm thick inert, metallic chromium (Cr(0), zero-valent chromium) capping layer using an ion beam sputter system. Chromium-capped ZVI films are stored in a glove box filled with high-purity argon at atmospheric pressure, for up to several days until the next stage of the experimental process. Chromium was selected as the capping agent for APT specimen preparation because (1) the mass-to-charge-state ratios of its isotopes do not overlap with that of zinc and sodium which are potential contaminants of the ZVI film and (2) chromium has a field-evaporation electric field closely matched with that of iron, thus allowing for smooth and controlled evaporation during the APT experiment when transitioning from the capping material into the sample surface.

**B. X-ray Photoelectron Spectroscopy (XPS).** XPS depth profile measurements were carried out using a Thermoscientific ESCALAB 250 Xii available at the NUANCE center at Northwestern University. The instrument is calibrated to give a binding energy of 83.96 eV for the Au $4f_{7/2}$ line and uses a K$\alpha$ radiation from a monochromatic aluminum source. A flood gun is used to compensate for surface charging by the ejection of low energy $Ar^+$ ions and electrons. To prevent the reduction of Fe(III) to Fe(II), we use the 2 mm raster size 2 keV etching mode ion ($Ar^+$) gun at mid current.[41] Etching time is set to 30 seconds. We note that these XPS conditions were chosen based on our previous study.[22]

**C. Focused Ion Beam Milling.** As briefly discussed above, APT involves field evaporation of specimen sample shapes that can produce an electric field on the order of $10^{10}$ V/nm in an ultrahigh vacuum chamber. Since the electric field produced is inversely proportional to the radius of curvature of the specimen, ZVI films along with the associated inert capping layer must be shaped into sharp conical tips. Here, focused ion beam (FIB) milling is the method employed to prepare conical shaped ZVI "tips". Fabrication details for the conically shaped ZVI tips (see Fig. 1A for the final shape of tips used in APT analysis) from inert capped ZVI films on silicon wafer substrates are provided in the Supporting Information (SI). Further details about FIB fabrication for atom probe tomography experiments can be found in Miller et al.[42]

**D. APT Analysis.** To evaluate the inherent features of nano ZVI films that could potentially explain the passivation of PRBs, we employ APT to obtain 3D images, and chemical composition of E-beam PVD prepared ZVI films. We briefly describe the atom probe process in the SI. Further descriptions of APT analysis and techniques are available in the literature.[37,43-44] For data collection, we use a LEAP4000XSi atom-probe tomograph manufactured by CAMECA.[43,45] The sharpened tips are cryogenically cooled inside the LEAP instrument to a temperature of about 44 K. Atoms from the surface of the sharp tip are field-evaporated by applying a high voltage in the range of 3-6 kV. Field evaporation is assisted by ultraviolet laser pulses with 30 pJ pulse energy and a 250 kHz pulse repetition rate.

The identification of elemental peaks in the APT mass spectrum is straightforward by comparison of the peaks at the respective mass-to-charge (m/z) ratios with known isotopic abundance patterns (see Fig. 1B). APT reconstruction and data analysis is performed utilizing the

IVAS 3.6.14 software package.

## III. Results.

**A. APT Reveals Dendritic Growth of Oxide Overlayer into Bulk ZVI.** Figure 1B shows the

APT mass spectrum of a ZVI tip specimen after most of the inert chromium cap had been field-evaporated. The mass spectrum contains peaks of the following ionic and molecular-ionic species from the ZVI film: $Fe^+$, $Fe^{2+}$, $FeH^+$, $FeO^+$, $Fe_2O^{2+}$, $FeO_2^+$, and $FeO_2^{2+}$. Additionally, m/z

peaks detected from residuals of the inert chromium capping layer include Cr, CrO, $CrO_2$, and

$CrO_3$. We caution that the ionic charge state (here 1+ or 2+) of the ions and molecular ions in

APT is a consequence of the field-ionization process during field-evaporation and not necessarily

related to the valency of the atom in its original chemical bonding environment. Chemical

valency in the ZVI film and its oxide overlayer is already known within the context of the XPS

results from our previous work.[22] For the initial overall reconstruction that includes the

chromium cap, the peaks at m/z = 27 and 54 [which are isobaric overlaps for both [54]Fe and [54]Cr

isotopes] are attributed to chromium. The peaks of the 54 isotopes for the molecular oxide

species of both Cr and Fe are treated analogously. Once the chromium capping is removed from

the dataset via thresholding with an isoconcentration surface at a level of 50 at.% Fe, the peaks at

m/z = 27 and 54 (and analogously for the molecular oxide species) are reassigned to Fe, since the

contribution of Cr to these peaks now is negligible. This is also confirmed by the height of the

peaks at m/z = 54 and 27 relative to the peaks of the other Fe isotopes, which, after removing the

Cr capping layer from the dataset, fit the isotopic abundance distribution for iron well (see Fig.

1B).

No evidence of m/z peaks matching the known natural isotopic abundance patterns for Zn (m/z = 32-35 or 64-70) and Na (m/z = 23), which are potential contaminants (through hydroxide-formation) of the ZVI films given their presence in the 99.95% purity iron boule used in preparing the iron nanofilm, are observed. This high level of purity is attributed to the E-beam evaporation method described in our prior work,[22] in which evaporation rates are set high enough to exceed the iron melting point of our 99.95% purity iron boules (~1540 °C). Our method circumvents the need for iron sources having a purity of at least 5N (99.999% Fe), which are considerably more expensive and difficult to obtain[46] and handle[46-47] than the commonly available 99.95% purity iron boules used here.

3D-reconstruction of the ZVI tips including the inert Cr capping shows three defined layers (see Fig. 2). The first layer from the top is comprised of Cr(0) and chromium-oxide ionic species in the inert capping layer. In the second layer, we find oxygen species ($O^+$ and $O_2^+$) and iron-oxygen molecular-ion species ($Fe_2O^{2+}$, $FeO^+$, $FeO_2^+$). The most abundant ionic species present in the third layer are $Fe^+$, $Fe^{2+}$, and $FeH^+$. Eliminating the inert Cr capping layer from the 3D reconstruction dataset (Fig. 3A) clearly reveals the presence of the iron-oxygen molecular-ion species in a ~5 nm thick overlayer over the ZVI film, consistent with the presence of the iron oxide overlayer we reported earlier, albeit based on the spatially-averaged results from XPS.[22]

A proxigram concentration profile[48-49] created from the full 3D APT reconstruction (see Fig. 3B) with an 50 atomic % isoconcentration surface of atomic Fe shows that the Cr concentration due to the capping layer is anticorrelated with the Fe concentration. The surface of the ZVI film is covered with an approximately 5 nm thick oxide layer, identified as the oxide nano-overlayer of the ZVI nanofilm, just like what is seen in Fig. 3A. XPS can resolve the valency of Fe and at the same time obtain an average oxygen concentration (Fig. 3C) with some depth resolution (Fig. 3B). Indeed, the presence of the iron oxide nano-overlayer film and the spatially averaged thickness thereof determined from the APT dataset agree with the concentration-depth profile obtained by XPS analysis (Fig. 3C).[22]

Further analysis of the interface between the iron oxide nano-overlayer and the zero-valent iron nanofilm reveals the presence of dendritic features extending from the iron oxide nano-overlayer. Fig. 4 and Fig. 5A display isoconcentration surfaces to reveal the morphology of the oxygen-rich regions growing into the bulk ZVI. Dendritic features grow ~10 nm into the Fe(0) ZVI bulk layer and are ~5 nm wide. Figure 5B shows a contour plot of the oxygen concentration in a 6 nm thin slice cut in the direction of the oxide dendrites. The graph reveals the oxygen distribution in the overlayer and the oxide dendrites. Similar features were observed in a second dataset (Fig. S3). The results indicate a dense distribution of closely spaced dendrites extending from the oxide nano-overlayer into the ZVI nanofilm. The dendrite morphology is probably related to the growth kinetics of the oxide overlayer film, which is beyond the scope of this present work. Currently, the limited field-of-view available in the APT reconstructions, which reveal only a small number of dendrites in a given graphical 2-dimensional slice, prevents us from determining the large-scale distribution of dendrite features within the ZVI films, but analysis is ongoing.

**B. Oxide Overlayer is Stable in Air over Months.** The overall structure of the ZVI films, which includes the ~ 5 nm oxide layer over the ~15 nm Fe(0) bulk, remains intact after exposure to ambient conditions for three months as supported by XPS depth profiling (Fig. 6). At the surface, both Fe(0) and Fe(III) peaks are observed. After ion-beam etching for 30 seconds (depth of ~5-10 nm) and 60 seconds (~10-15 nm), only Fe(0) peaks are observed. This result suggests the growth of the iron oxide nano-overlayer halts or continues at a negligibly slow rate once a depth of approximately 10 nm has been reached. This 10 nm thick overlayer grows initially at a rate that we are planning to determine in future studies, but we suspect the oxide overlayer to be formed during the first few minutes of exposing the ZVI film to air. XPS depth profiling results of newly prepared and older (~ 3 months) ZVI films do not reveal the presence of Fe(II) within the ZVI nanofilms. The absence of an intermediate Fe(II) layer raises the question as to whether the Fe(0) is directly oxidized to Fe(III) during the first few minutes of exposure to atmospheric conditions for freshly made ZVI films. As stated in the Section IIIA, freshly prepared ZVI films are usually exposed to atmospheric conditions for ~ 30 minutes in our studies before further experimental procedures are performed. Moving forward, and with the modification of our existing experimental setup, it will be of interest to see if reducing this exposure time to a period of seconds or minutes can reveal more about the kinetics and intermediate stages of the initial growth of the ~ 5 nm oxide overlayer, as our current experimental set-up does at present not allow for an efficient reduction in the exposure time. Additional experiments with varying levels of oxygen and/or relative humidity are being planned as well.

**IV. Discussion.** Our XPS and APT results suggest that the oxide layer and its dendritic features form as soon as the ZVI films are removed from the UHV chamber used for E-beam PVD and exposed to atmospheric ambient conditions, within a few minutes. The isoconcentration surfaces in the APT 3D reconstructions (Figs. 3-6) give an indication of the curvy nature of the dendritic iron oxide:iron interface. Compositional profiling with the APT proxigram (Fig. 3A) considers the curvy shape of the oxide interface since the proxigram analyzes the composition with distance from that isoconcentration surface. The oxygen concentration estimates provided by APT, while spatially resolved on a nano-scale, may, however, be too low due to the limitations with oxygen detection in APT. XPS, on the other hand, gives us a better oxygen quantification (Fig. 3C), but with no lateral resolution due to its averaging over the entire surface sampled. The difference in the atomic oxygen percentages seen from the APT proxigram profile and the XPS depth profile (Figs. 3A & 3B) can be attributed to the fact that APT has a limited capability with oxygen quantification due to the formation of various neutral oxygen species during field-evaporation in APT.[39-40] XPS, on the other hand, does better at quantifying the amount of oxygen present within the ZVI films at the detriment of limited spatial resolution of the oxygen profile within the films due to the loss of intermediate information in the course of sputtering and the peak fitting averaging procedure.[22,41] We note that the pattern and distribution of the dendritic features are not indicative of any grain boundary pattern or structure, as far as we can tell.

In our prior work, we showed that after the initial growth phase, the ZVI oxide overlayer thickness (~ 5 nm) remains invariant with the duration of exposure to ambient conditions.[22] The dendritic features of the oxide overlayer, as revealed by APT analysis, is evidence of penetrative oxidation propagation within the nano-sized ZVI films. The iron bulk is known from grazing angle x-ray diffraction measurements to be at least partly crystalline in nature whereas the oxide overlayer appears to be a mixture of different iron oxides including magnetite from prior Raman studies.[22] The films are known to be devoid of Na and Zn which readily form oxide and hydroxides that can destabilize the films.[22] Therefore, we argue that the dendrites form through an electrochemistry process, specifically, direct redox reactions involving the Fe(0) present within the films.[1,13]

Penetrative oxidation by oxygen diffusion has been reported to occur in bulk alloys under extreme conditions such as exposure to steam, hot pressurized water and stress corrosion cracking,[15-16] but not for nanofilms of relatively pure metals under ambient conditions. For instance, APT studies on Ni-30Cr alloys exposed to pressurized hot water indicate the presence of filaments of oxide with a Cr/Ni oxide surrounding platelets of $Cr_2O_3$.[16] Likewise, oxygen diffusion with dendrite-like precursors preceding the oxidation front has been shown to occur ahead of crack tips in stress-corrosion cracked alloy X-750 in boiling water reactor environments.[15] Recently, Li et al. studied the topmost atomic layers of iridium oxide, an electrocatalyst used in water splitting, chlor-electrolysis, or fuel cells, using APT, XPS, and transmission electron microscopy.[50] They reported the formation of a thin (<5 nm) metastable oxide with an Ir to O ratio of 1:1 that forms during the first four hours of anodic oxidation of an iridium film. Though Li et al. did not report a dendritic oxidation growth of the iridium oxide and related oxide propagation to grain boundaries in a nano-crystalline structure, a careful look at the published APT reconstructions reveals evidence of a dendritic oxide growth morphology within the iridium film.[50] Iridium is a noble metal stable under ambient conditions that can be oxidized only through electrochemical means.[50] Since iridium does not readily oxidize under atmospheric conditions like iron, these authors were able to track a change of the stoichiometric ratio between Ir and O over time. However, the rapid oxidation of iron prevents us from distinguishing different stages for stoichiometric growth of the O and Fe, from when the freshly deposited ZVI films are removed from E-beam PVD UHV chamber and analyzed with XPS within our current experimental time constraint (~ 30 min).

A complete eventual understanding the formation kinetics of the dendritic oxide overlayer holds the promise of opening a path towards designs that can help overcome passivation and corrosion limitations in ZVI nanofilms. In essence, if the formation of the dendritic oxide layer is progressive but slow, ZVI nanofilms could be prepared in a manner that continually exposes fresh Fe(0) atoms to exchange electrons in the redox reaction with heavy metal toxic contaminants even during the dendritic oxide growth process. Nonetheless, if a solid continuous oxide layer proves terminal in terms of material function, ZVI nanofilms growth could be modified to contain a local surface structure that allows for the exposure of more Fe(0) atoms to react with toxic metal contaminant at the surface, for instance, in applications aimed at improved PRBs. Concrete designs for other applications, including corrosion inhibition of iron-bearing materials, films, and their surfaces would require computer-guided modeling, which is beyond the scope of our present work. Yet, our present results set the stage for investigating the atmospheric corrosion process in the limit of having low impurities or limited quantities of heavy metals in contact with ZVI films. Subsequent work will focus on how dendrite formation in ZVI films can be promoted or mitigated and whether the dendrites play a role in surface passivation or corrosion. Finally, we envision the APT images presented here are useful for scientists and engineers aiming to build particle-in-3D-box models with spatially alternating potentials on one side of the box, just like the APT image shows it should be. It will also be interesting to see if the dendrites we identify in this study can be further modified to promote or disrupt the charge carrier motion in the iron nanofilms or steer it.

**V. Conclusions.** Atom probe tomography (APT) analysis of zero-valent iron (ZVI) nanofilms with potential use in a variety of environmental, optoelectronic, chemical, propulsion, and energy technologies reveals the presence of dendritic iron oxide features that extend from a surface oxide overlayer into the ZVI bulk. The dendrites are observed by APT to be in the 5 nm x 10 nm size range, form quickly under natural atmospheric conditions and their growth into the films is, within the limit of our three-month long study, self-limiting (i.e. their initial growth appears to quickly discontinue). We hypothesize that the presence of oxidized iron dendrites over ZVI is likely to be related to the passivation processes that currently hampers the long-term use of PRBs for enviromnmental remediation. Re-engineering based on a detailed analysis may provide a pathway to impede, halt, or even reverse the passivation process. The atomistic views also shed first light on the atmospheric corrosion process of Fe(0)-bearing engineered materials and their surfaces in the limit of low bulk impurities.

**Supporting Information.** Contains a description of the FIB fabrication process, second dataset of APT images showing dendritic features of another ZVI tip, and a graph of the APT ionic concentration profile of iron-oxygen species in a typical 20 nm ZVI film. Supporting information is available free of charge on the ACS Publications website.

**VI. Acknowledgments.** This work was supported by the US National Science Foundation (NSF). FMG gratefully acknowledges NSF award number CHE-1464916 and support from a Friedrich Wilhelm Bessel Prize from the Alexander von Humboldt Foundation. The XPS work was performed in the Keck-II facility of NU*ANCE* Center at Northwestern University. The NU*ANCE* Center is supported by the International Institute for Nanotechnology, MRSEC (NSF DMR-1121262), the Keck Foundation, the State of Illinois, and Northwestern University. This work made use of the EPIC facility (NU*ANCE* Center-Northwestern University), which has received support from the MRSEC program (NSF DMR-1121262) at the Materials Research Center; the International Institute for Nanotechnology (IIN); and the State of Illinois, through the IIN. Atom-probe tomography was performed at the Northwestern University Center for Atom-Probe Tomography (NUCAPT). The LEAP tomograph at NUCAPT was purchased and upgraded with grants from the NSF-MRI (DMR-0420532) and ONR-DURIP (N00014-0400798, N00014-0610539, N00014-0910781, N00014-1712870) programs. NUCAPT received support from the MRSEC program (NSF DMR-1720139) at the Materials Research Center, the SHyNE Resource (NSF ECCS-1542205), and the Initiative for Sustainability and Energy (ISEN) at Northwestern University.

**Author Contributions.** MB performed the experiments. MB, DI, and FMG analyzed the data.

The manuscript was written with substantial contributions from MB, DI, and FMG.

**Notes.** The authors declare no competing financial interests.

## References

1.      Fu, F.; Dionysiou, D. D.; Liu, H., The Use of Zero-Valent Iron for Groundwater Remediation and Wastewater Treatment: A Review. *J. Hazard. Mater.* **2014**, *267*, 194-205.

2.      Homann, K.; Freund, H. J., $N_2$ Adsorption and Dissociation on Thin Iron Films on W(110). *Surf. Sci.* **1995**, *327*, 216-224.

3.      Thielemann, P.; Brandt, U., Magentooptical Studies on Thin Iron Films. *Appl. Phys. A* **1982**, *28*, 53-58.

4.      Gerhard, L.;Yamada, T. K.; Balashov, T.; Takacs, A. F.; Wesselink, R. J. H.; Daene, M.; Fechner, M.; Ostanin, S.; Ernst, A.; Mertig, I.; Wulfhekel, W., Magnetoelectric Coupling at Metal Surfaces. *Nat. Nanotechnol.* **2010**, *5*, 792-797.

5.      Okoshi, M.; Awaihara, Y.; Yamashita, T.; Inoue, N., $F_2$-Laser-Induced Surface Modification of Iron Thin Films to Obtain Corrosion Resistance. *Jpn. J. Appl. Phys.* **2014**, *53*, 022702.

6.      Eary, L. E.; Rai, D., Kinetics of Chromate Reduction by Ferrous-Ions Derived from Hematite and Biotite at 25-Degrees-C. *Am. J. Sci.* **1989**, *289*, 180-213.

7.      Kendelewicz, T.; Liu, P.; Doyle, C. S.; Brown Jr, G. E.; Nelson, E. J.; Chambers, S. A., X-Ray Absorption and Photoemission Study of the Adsorption of Aqueous Cr(VI) on Single Crystal Hematite and Magnetite Surfaces. *Surf. Sci.* **1999**, *424*, 219-231.

8.      Ilton, E. S.; Veblen, D. R., Chromium Sorption by Phlogopite and Biotite in Acidic Solutions at 25°C: Insights from X-Ray Photoelectron Spectroscopy and Electron Microscopy. *Geochim. Cosmochim. Acta* **1994**, *58*, 2777-2788.

9.      Bidoglio, G.; Gibson, P. N.; O'Gorman, M.; Roberts, K. J., X-Ray Absorption Spectroscopy Investigation of Surface Redox Transformations of Thallium and Chromium on Colloidal Mineral Oxides. *Geochim. Cosmochim. Acta* **1993**, *57*, 2389-2394.

10.     Blowes, D. W.; Ptacek, C. J.; Jambor, J. L., In-Situ Remediation of Cr(VI)-Contaminated Groundwater Using Permeable Reactive Walls: Laboratory Studies. *Environ. Sci. Technol.* **1997**, *31*, 3348-3357.

11.     Mackay, D. M.; Cherry, J. A., Groundwater Contamination: Pump-and-Treat Remediation. *Environ. Sci. Technol.* **1989**, *23*, 630-636.

12.     Troiano, J. M.; Jordan, D. S.; Hull, C. J.; Geiger, F. M., Interaction of Cr(III) and Cr(VI) with Hematite Studied by Second Harmonic Generation. *J. Phys. Chem. C* **2013**, *117*, 5164-5171.

13.     Qiu, S. R.; Lai, H. F.; Roberson, M. J.; Hunt, M. L.; Amrhein, C.; Giancarlo, L. C.; Flynn, G. W.; Yarmoff, J. A., Removal of Contaminants from Aqueous Solution by Reaction with Iron Surfaces. *Langmuir* **2000**, *16*, 2230-2236.

14.     Davenport, A. J.; Oblonsky, L. J.; Ryan, M. P.; Toney, M. F., The Structure of the Passive Film That Forms on Iron in Aqueous Environments. *J. Electrochem. Soc.* **2000**, *147*, 2162-2173.

15.     Gibbs, J. P.; Ballinger, R. G.; Jackson, J. H.; Isheim, D.; Hänninen, H., Stress Corrosion Cracking and Crack Tip Characterization of Alloy X-750 in Boiling Water Reactor Environments. In *15th International Conference on Environmental Degradation of Materials in Nuclear Power Systems-Water Reactors*, John Wiley & Sons, Inc.: **2012**; pp 745-760.

16.     Olszta, M. T., L.; Bruemmer, S. , Resolving Nanostructures in Complex Penetrative Oxidation for Ni-30Cr Alloys Exposed to High-Temperature Water using APT and TEM. *Microsc. Microanal.* **2011**, *17*, 734-735.

17.     Eggleston, C. M.; Stumm, W., Scanning Tunneling Microscopy of Cr(III) Chemisorbed on α-Fe$_2$O$_3$ (001) Surfaces from Aqueous Solution: Direct Observation of Surface Mobility and Clustering. *Geochim. Cosmochim. Acta* **1993**, *57*, 4843-4850.

18.     Charlet, L.; Manceau, A. A., X-Ray Absorption Spectroscopic Study of the Sorption of Cr(III) at the Oxide-Water Interface: Ii. Adsorption, Coprecipitation, and Surface Precipitation on Hydrous Ferric Oxide. *J. Colloid Interface Sci.* **1992**, *148*, 443-458.

19.     Hori, M.; Shozugawa, K.; Matsuo, M., Reduction Process of Cr(VI) by Fe(II) and Humic Acid Analyzed Using High Time Resolution XAFS Analysis. *J. Hazard. Mater.* **2015**, *285*, 140-147.

20.     Taylor, S. D.; Liu, J.; Arey, B. W.; Schreiber, D. K.; Perea, D. E.; Rosso, K. M., Resolving Iron(II) Sorption and Oxidative Growth on Hematite (001) Using Atom Probe Tomography. *J. Phys. Chem. C* **2018**, *122*, 3903-3914.

21.     Gault, B.; Menand, A.; de Geuser, F.; Deconihout, B.; Danoix, R., Investigation of an Oxide Layer by Femtosecond-Laser-Assisted Atom Probe Tomography. *Appl. Phys. Lett.* **2006**, *88*, 3.

22.     Faurie-Wisniewski, D.; Geiger, F. M., Synthesis and Characterization of Chemically Pure Nanometer-Thin Zero-Valent Iron Films and Their Surfaces. *J. Phys. Chem. C* **2014**, *118*, 23256-23263.

23.     Geiger, F. M.; Wisniewski, D. F. Chemically Pure Zero-Valent Iron Nanofilms from a Low-Purity Iron Source. US Patent Number 9738966, 22 August, **2017**.

24.     Stevens, S. M.; Cubillas, P.; Jansson, K.; Terasaki, O.; Anderson, M. W., Nanoscale Electron Beam Damage Studied by Atomic Force Microscopy. *J. Phys. Chem. C* **2009**, *113*, 18441-18443.

25.     Jung, H.-G.; Kim, H.-S.; Park, J.-B.; Oh, I.-H.; Hassoun, J.; Yoon, C. S.; Scrosati, B.; Sun, Y.-K., A Transmission Electron Microscopy Study of the Electrochemical Process of Lithium–Oxygen Cells. *Nano Lett.* **2012**, *12*, 4333-4335.

26.     MacLaren, I.; Ramasse, Q. M., Aberration-Corrected Scanning Transmission Electron Microscopy for Atomic-Resolution Studies of Functional Oxides. *Int. Mater. Rev.* **2014**, *59*, 115-131.

27.     Ke, X.; Bittencourt, C.; Van Tendeloo, G., Possibilities and Limitations of Advanced Transmission Electron Microscopy for Carbon-Based Nanomaterials. *Beilstein J. Nanotechnol.* **2015**, *6*, 1541-1557.

28.     Sarangi, R., X-Ray Absorption near-Edge Spectroscopy in Bioinorganic Chemistry: Application to M–O$_2$ Systems. *Coord. Chem. Rev.* **2013**, *257*, 459-472.

29.     Latif, C.; Negara, V. S. I.; Wongtepa, W.; Thamatkeng, P.; Zainuri, M.; Pratapa, S., Fe K-Edge X-Ray Absorption Near-Edge Spectroscopy (XANES) and X-Ray Diffraction (XRD) Analyses of Life Po-4 and Its Base Materials. *J. Phys: Conf. Ser.* **2018**, *985*, 012021.

30.     Schleiwies, J.; Schmitz, G.; Heitmann, S.; Hutten, A., Nanoanalysis of Co/Cu/Nife Thin Films by Tomographic Atom Probe. *Appl. Phys. Lett.* **2001**, *78*, 3439-3441.

31.     Jessner, P.; Goune, M.; Danoix, R.; Hannoyer, B.; Danoix, F., Atom Probe Tomography Evidence of Nitrogen Excess in the Matrix of Nitrided Fe-Cr. *Philos. Mag. Lett.* **2010**, *90*, 793-800.

32.     Larson, D. J., Atom Probe Characterization of Nanomagnetic Materials. *Thin Solid Films* **2006**, *505*, 16-21.

33.    Kuhlman, K. R.; Martens, R. L.; Kelly, T. F.; Evans, N. D.; Miller, M. K., Fabrication of Specimens of Metamorphic Magnetite Crystals for Field Ion Microscopy and Atom Probe Microanalysis. *Ultramicroscopy* **2001**, *89*, 169-176.

34.    Pereloma, E. V.; Stohr, R. A.; Miller, M. K.; Ringer, S. P., Observation of Precipitation Evolution in Fe-Ni-Mn-Ti-Al Maraging Steel by Atom Probe Tomography. *Metall. Mater. Trans. A* **2009**, *40A*, 3069-3075.

35.    Larson, D. J.; Cerezo, A.; Juraszek, J.; Hono, K.; Schmitz, G., Atom-Probe Tomographic Studies of Thin Films and Multilayers. *MRS Bull.* **2009**, *34*, 732-737.

36.    Marceau, R. K. W.; Ceguerra, A. V.; Breen, A. J.; Raabe, D.; Ringer, S. P., Quantitative Chemical-Structure Evaluation Using Atom Probe Tomography: Short-Range Order Analysis of Fe-Al. *Ultramicroscopy* **2015**, *157*, 12-20.

37.    Kelly, T. F.; Miller, M. K., Invited Review Article: Atom Probe Tomography. *Rev. Sci. Instrum.* **2007**, *78*, 20.

38.    Kelly, T. F.; Larson, D. J.; Thompson, K.; Alvis, R. L.; Bunton, J. H.; Olson, J. D.; Gorman, B. P., Atom Probe Tomography of Electronic Materials. In *Annu. Rev. Mater. Res.*, Annual Reviews: Palo Alto, **2007**; Vol. 37, pp 681-727.

39.    Bachhav, M.; Danoix, F.; Hannoyer, B.; Bassat, J. M.; Danoix, R., Investigation of O-18 Enriched Hematite ($\alpha$-Fe$_2$O$_3$) by Laser Assisted Atom Probe Tomography. *Int. J. Mass Spectrom.* **2013**, *335*, 57-60.

40.    Devaraj, A.; Colby, R.; Hess, W. P.; Perea, D. E.; Thevuthasan, S., Role of Photoexcitation and Field Ionization in the Measurement of Accurate Oxide Stoichiometry by Laser-Assisted Atom Probe Tomography. *J. Phys. Chem. Lett.* **2013**, *4*, 993-998.

41.    Yin, L. I.; Ghose, S.; Adler, I., X-Ray Photoelectron Spectroscopic Studies of Valence States Produced by Ion-Sputtering Reduction. *Appl. Spectrosc.* **1972**, *26*, 355-357.

42.    Miller, M. K.; Russell, K. F.; Thompson, K.; Alvis, R.; Larson, D. J., Review of Atom Probe Fib-Based Specimen Preparation Methods. *Microsc. Microanal.* **2007**, *13*, 428-436.

43.    Larson David J., P. T. J., Ulfig Robert M., Geiser Brian P., Kelly Thomas F., *Local Electrode Atom Probe Tomography*; Springer Science+Business Media: New York, **2013**, p 328.

44.    Gault, B., Moody, M.P., Cairney, J.M., Ringer, S.P., *Atom Probe Microscopy*; Springer Series in Materials Science, **2012**.

45.    Adusumilli, P.; Murray, C. E.; Lauhon, L. J.; Avayu, O.; Rosenwaks, Y.; Seidman, D. N., Three-Dimensional Atom-Probe Tomographic Studies of Nickel Monosilicide/Silicon Interfaces on a Subnanometer Scale. In *Advanced Gate Stack, Source/Drain, and Channel Engineering for Si-Based Cmos 5: New Materials, Processes, and Equipment*, Narayanan, V.; Roozeboom, F.; Kwong, D. L.; Iwai, H.; Gusev, E. P.; Timans, P. J., Eds. Electrochemical Soc Inc: Pennington, **2009**; Vol. 19, pp 303-314.

46.    Okoshi, M.; Awaihara, Y.; Yamashita, T.; Inoue, N., F$_2$-Laser-Induced Surface Modification of Iron Thin Films to Obtain Corrosion Resistance. *Jpn. J. Appl. Phys.* **2014**, *53*.

47.    Nagano, M.; Hayashi, Y.; Ohtani, N.; Isshiki, M.; Igaki, K., Hydrogen Diffusivity in High-Purity Alpha-Iron. *Scr. Metall.* **1982**, *16*, 973-976.

48.    Hellman, O. C.; Vandenbroucke, J. A.; Rusing, J.; Isheim, D.; Seidman, D. N., Analysis of Three-Dimensional Atom-Probe Data by the Proximity Histogram. *Microsc. Microanal.* **2000**, *6*, 437-444.

49.    Hellman, O. C.; Rivage, J. B. d.; Seidman, D. N., Efficient Sampling for Three-Dimensional Atom Probe Microscopy Data. *Ultramicroscopy* **2003**, *95*, 199-205.

50.     Li, T.; Kasian, O.; Cherevko, S.; Zhang, S.; Geiger, S.; Scheu, C.; Felfer, P.; Raabe, D.; Gault, B.; Mayrhofer, K. J. J., Atomic-Scale Insights into Surface Species of Electrocatalysts in Three Dimensions. *Nat. Catal.* **2018**, *1*, 300-305.

**Figures and Captions.**

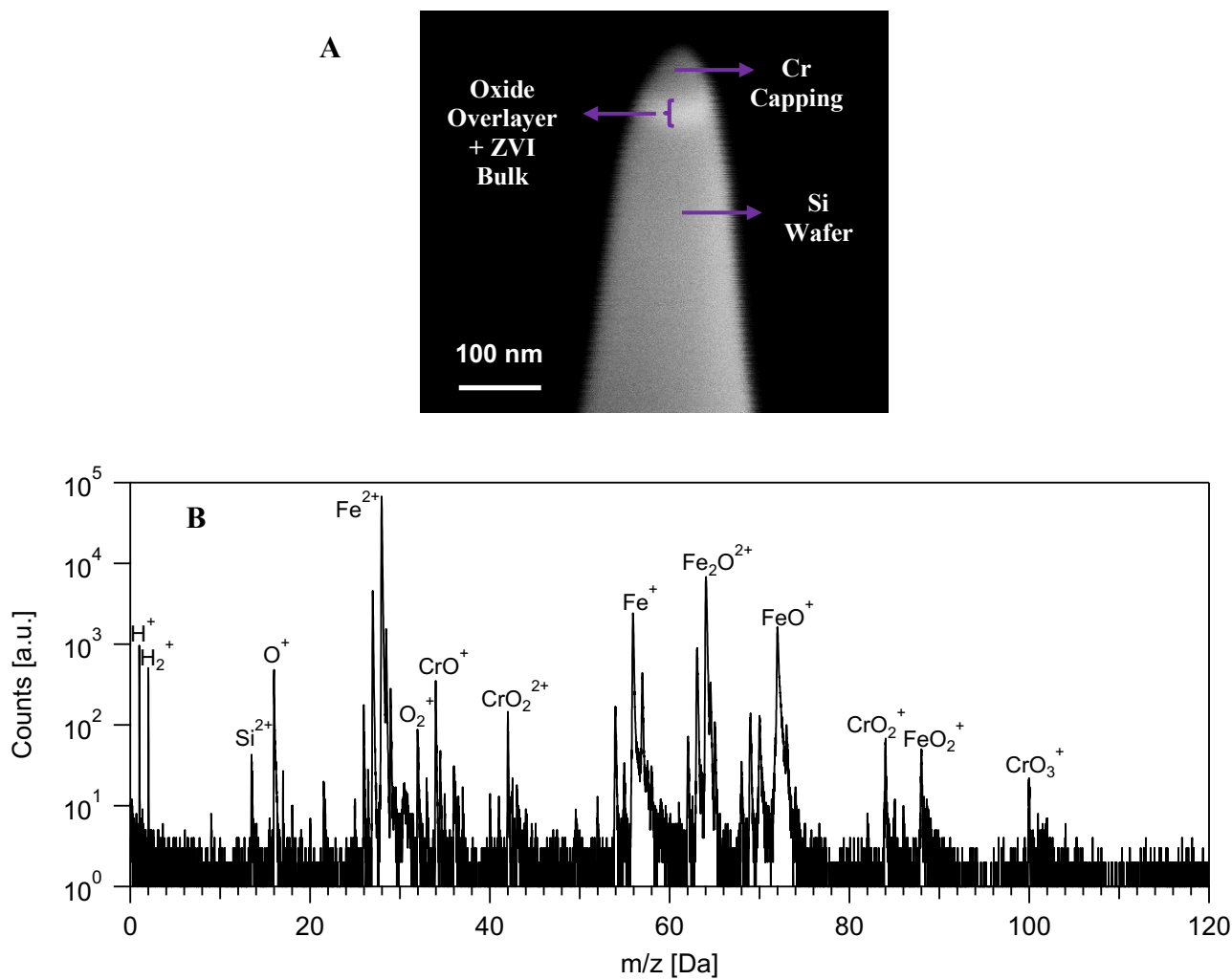

**Figure 1. (A)** Scanning electron microscopy image for the final shape of a needle-shaped Chromium capped ZVI tip on a silicon substrate/post ready for APT analysis [Image taken at 1.4nA and 5kV] **(B)** APT mass spectrum of a typical ZVI tip sample. The peaks with the highest natural isotopic abundance for each identified ionic or molecular-ionic species are labeled. Chromium species labeled here are remnants of the inert chromium cap at the ZVI interface after the chromium cap is separated from the ZVI layers following the APT reconstruction.

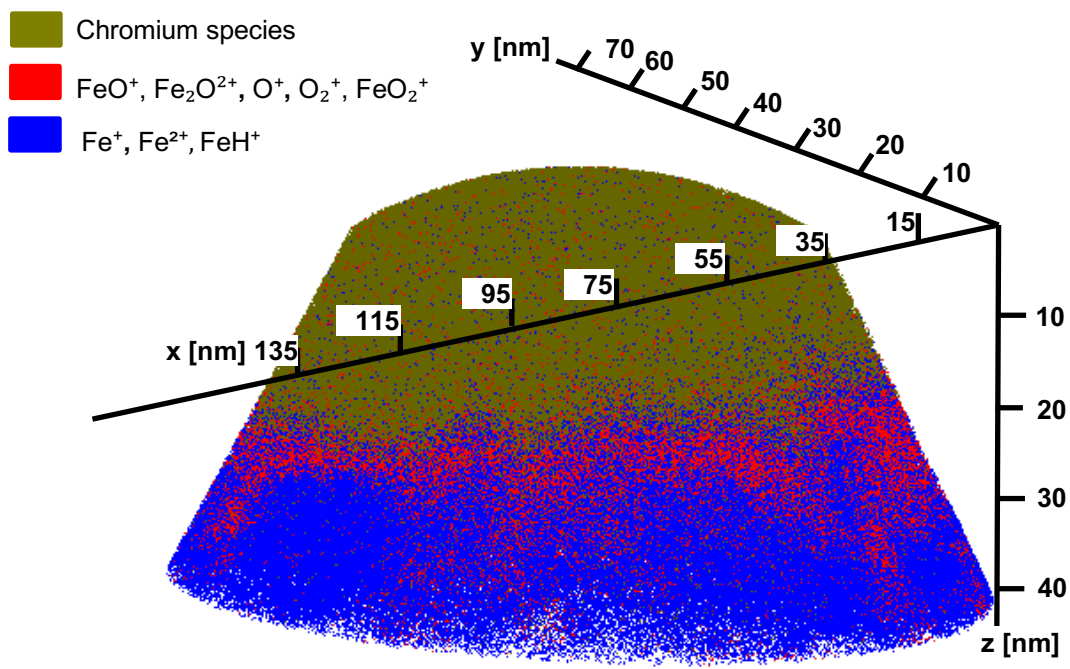

**Figure 2.** APT 3D reconstruction of a typical ZVI tip ([$Fe^+$, $Fe^{2+}$ and $FeH^+$]-blue, $FeO^+$, [$Fe_2O^{2+}$, $O_2^+$ and $O^+$]-red) with inert chromium (shown in olive green) capping.

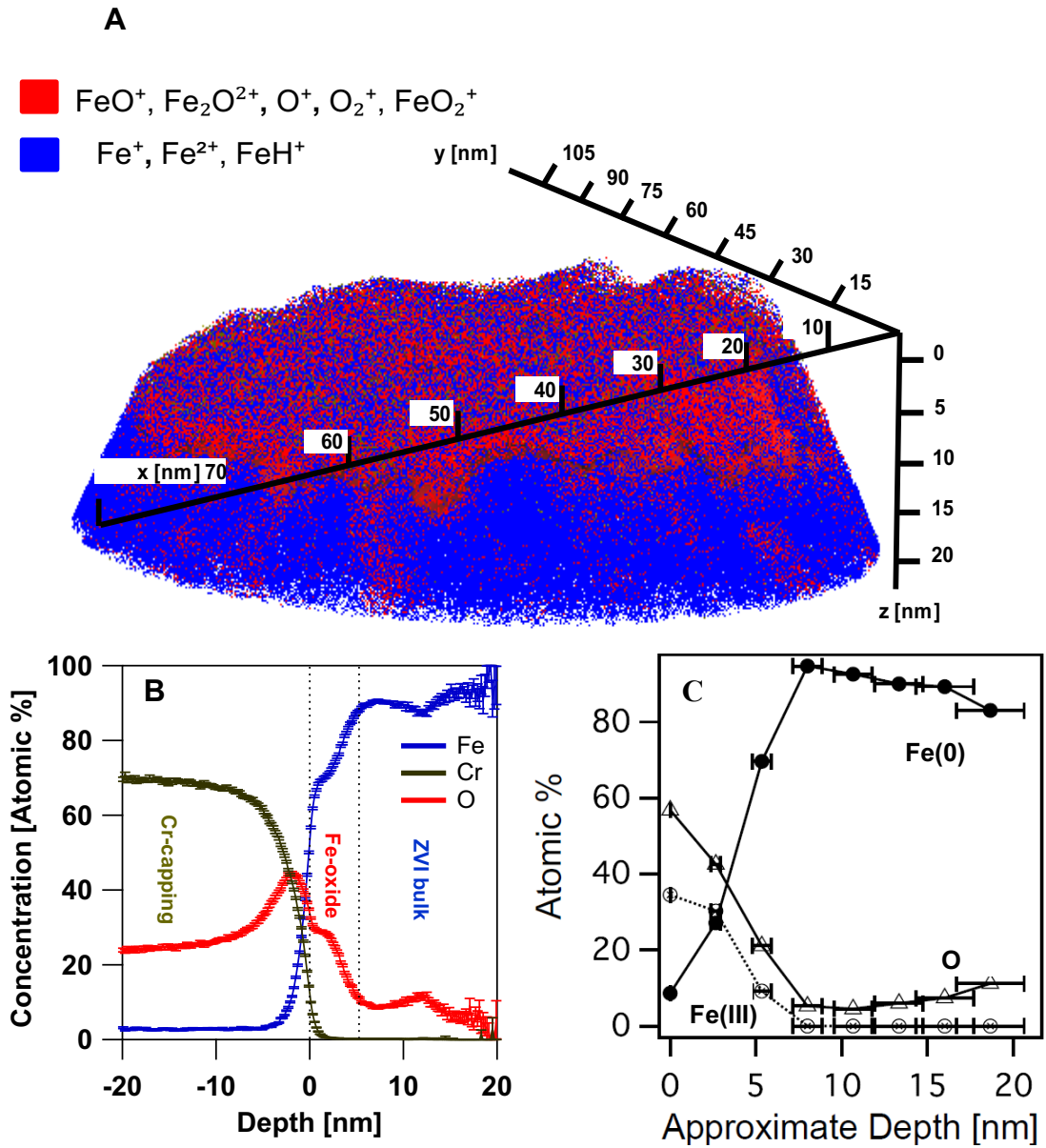

**Figure 3.** (**A**) 3D APT reconstruction of a typical ZVI tip, with the chromium capping layer removed by thresholding with a 50% isoconcentration surface of atomic iron species. An approximately ~5 nm thick layer of $FeO^+$, $Fe_2O^{2+}$, $FeO_2^+$, $O_2^+$, and $O^+$ (depicted in red) is visible on top of the ~ 10 nm iron [$Fe^+$, $Fe^{2+}$ and $FeH^+$- depicted in blue] bulk ZVI layer (**B**) Changes in atomic concentration (at.%) of Fe, Cr, O and Al with depth beginning with the Chromium capping at the left-hand side of the diagram. The interface of the chromium and ZVI layers is set to 0 nm. The oxidized iron layer includes $FeO^+$, $Fe_2O^{2+}$, $O_2^+$, and $O^+$. Dashed lines are used to delineate the three layers; inert Cr-capping, oxidized iron (Fe-Oxide), and ZVI bulk (Fe(0)). (**C**) Change in atomic % of Fe(0), Fe(III), and O as a function of depth of ~ 20 nm freshly prepared ZVI film from XPS analysis. A sputtering rate of 0.1778 nm/s was used for the depth estimation. The sputtering rate was determined by finding the ratio between the estimated thickness of the film from quartz crystal microbalance and ellipsometry measurements and the total sputter time taken to reach the Si substrate. Reproduced with permission from ref 24. Copyright 2014 American Chemical Society. Note that a freshly prepared ZVI film is exposed to atmospheric conditions for ~30 minutes before XPS is performed. For a colored representation of this Figure, please see to the web version of this article.

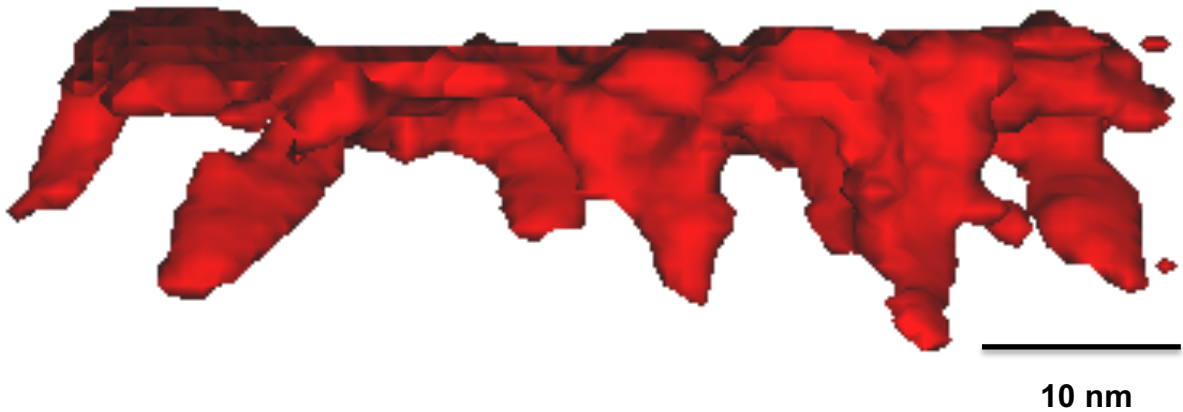

**Figure 4.** 1.4 Atomic % **i**soconcentration surface of $FeO^+$ + $Fe_2O^{2+}$ + $O^+$ + $O_2^+$ concentration (red) extracted from the 3D APT reconstruction of the ZVI tip in Fig. 5A, depicting the dendritic features (Confidence sigma level -10%). Image is projected in the xz plane.

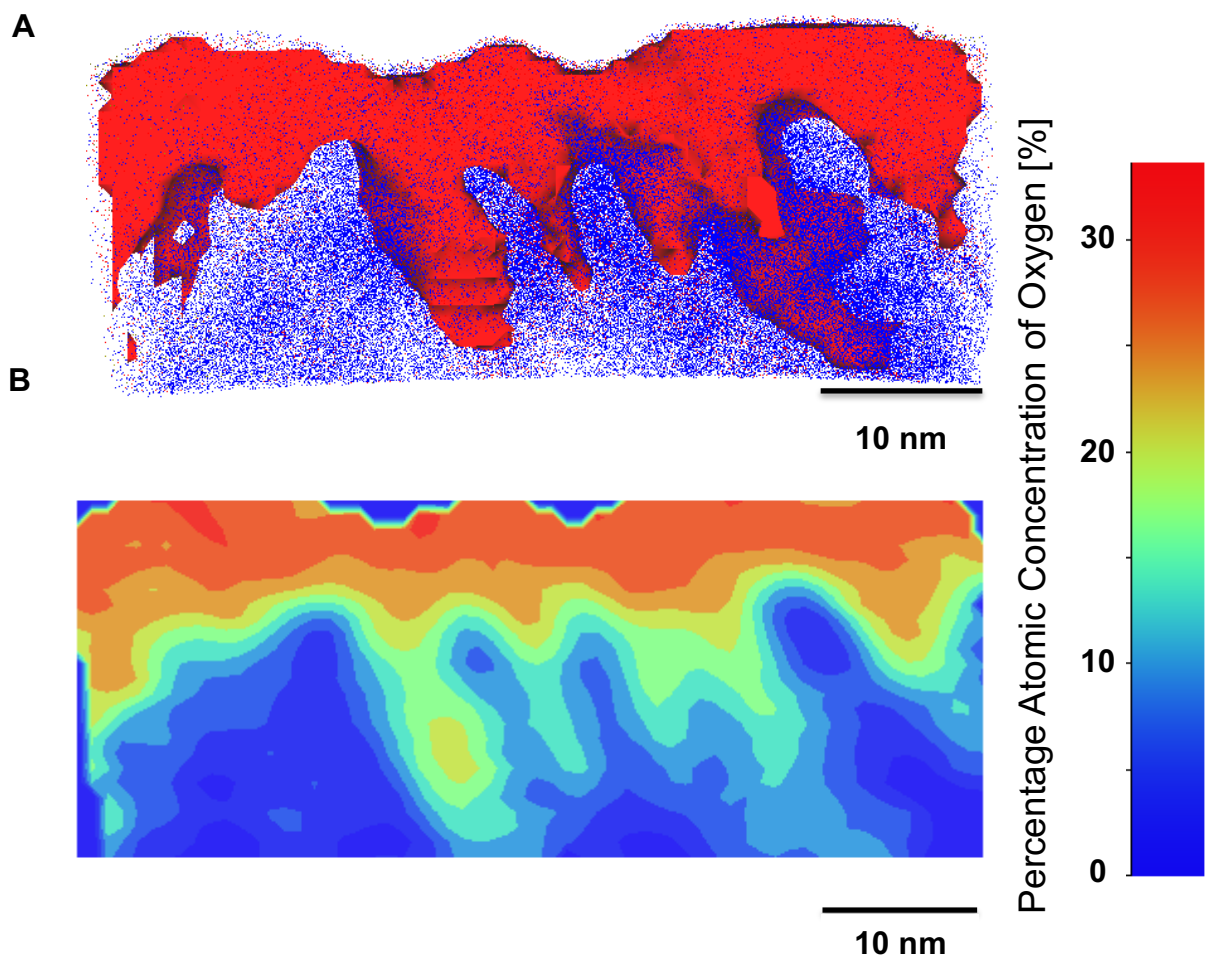

**Figure 5. (A)** 15 Atomic % isoconcentration surface of FeO$^+$ + Fe$_2$O$^{2+}$ + O$^+$ + O$_2^+$ concentration (red) of a cut portion of the above ZVI tip APT reconstruction shown in Fig. 3A, depicting the dendritic features extending from the oxide overlayer into the bulk ZVI film at the bottom. Blue dots represent Fe$^+$, Fe$^{2+}$ and FeH$^+$ ions, delineating the bulk of the ZVI film. The ZVI tip is sliced 6 nm in the x-direction and the image is projected in the yz plane. **(B)** 2D Contour plot of the oxygen concentration for Fig. 5A depicting the oxygen concentration gradient within the dendritic features. Color scale (see right) goes from blue to red with the red areas representing the highest concentration of FeO$^+$, Fe$_2$O$^{2+}$, O$^+$ and O$_2^+$. Blue represents zero oxygen content.

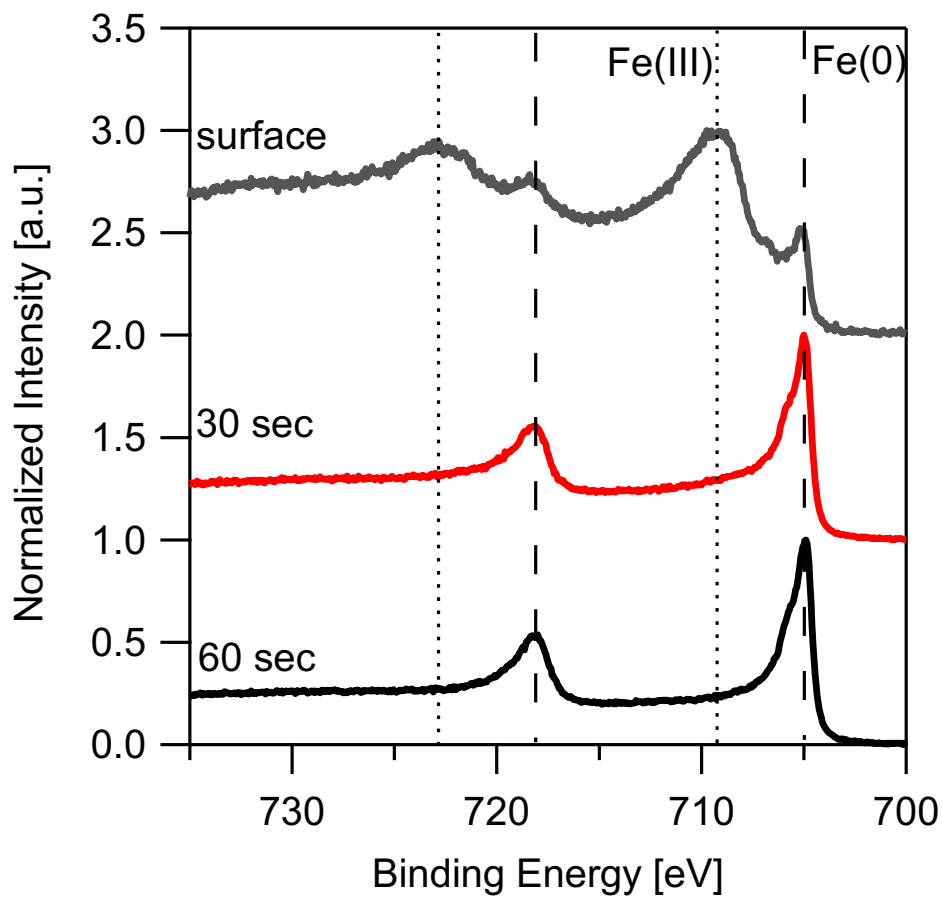

**Figure 6.** XPS (Fe 2p region) depth profile of a ~20 nm ZVI film that had been stored under ambient (air) conditions for three months. An etching time (see left of each intensity line) of 30 seconds corresponds to ~ 5-10 nm depth of ZVI material. Dashed lines represent peaks for Fe(0) and dotted lines represent peaks for Fe(III).

TOC Graphic

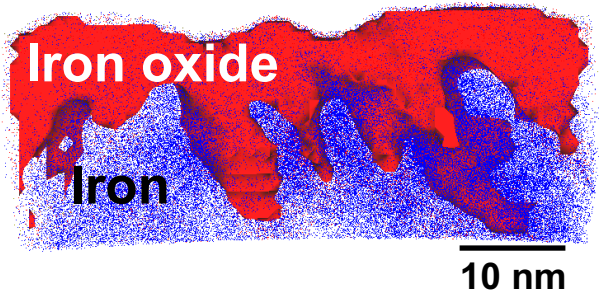

**Supporting Information for**

**Dendritic Oxide Growth in Zero-Valent Iron Nanofilms Revealed by Atom Probe Tomography**

Mavis D. Boamah,[1] Dieter Isheim[2], and Franz M. Geiger[1,*]

[1]Department of Chemistry, Northwestern University, Evanston, IL 60208, [2]Department of Material Science, and Center for Atom Probe Tomography, Northwestern University, Evanston, IL 60208, USA

*Corresponding author: geigerf@chem.northwestern.edu

**Focused Ion Beam Milling.** We used focused ion beam (FIB) milling to fabricate conical shaped ZVI "tips". The local electrode APT analytical method requires the use of conical shaped tips with end radius of curvature of ~50 nm and 100 μm height. The FIB process has two main stages. The first stage shown in Fig.1 involves making a wedge-shaped lift-out from a region of interest of the ZVI film with a protective capping (see Fig S1.A.). The wedge is then lifted out with a micromanipulator and a section ("blank") of it is mounted onto a ~150 μm tall silicon post on a micropost array (see Figs. S1.B and S1.C) by depositing platinum patches as seen in Fig.S1.D. The next stage of the FIB process (Fig. 2) involves sharpening of the blank into a conically-shaped sharp tip. The sharpening is done by first trimming (Figs. S2.A and S2.B) the broad sides of the blank into a pyramidal shape, followed by annular milling (Fig. S2.C) to achieve the desired conical shape and sharpness (Fig S2.D).

**Atom Probe Tomography Procedure.** A sample tip is mounted onto a nano-positioning stage and pointed towards a funnel-shaped local electrode.  With the help of two optical microscopes, the specimen tip is aligned roughly opposite to center of the local electrode's aperture. An electric field is then introduced on the specimen tip by applying a direct voltage which results in the initiation of ion-emission by field evaporation.  Using initial ion events from the inert chromium cap, a 30 pJ laser beam spot (pulse rate 250 kHz) is aligned on the specimen tip. The laser beam spot is centered on the apex of the specimen's tip and scanned continuously in the *xy* direction of a small scanning area during the evaporation process to allow for drift compensation. Once a sufficient signal is achieved and the *xy* scan is reproducible after three consecutive scans, a focus scan is performed. The detection rate is gradually increased in addition to voltage ramping as the tip is aligned. The aim is to obtain a steady rate of detected ion events.  The species that are emitted are ionized and projected radially onto a position sensitive 2D micro-channel plate (MCP) detector. The time-of-flight mass spectrometer allows for the detection of the mass-to-charge ratio of each ion based on the time it takes to reach the detector.

**Figures and Captions.**

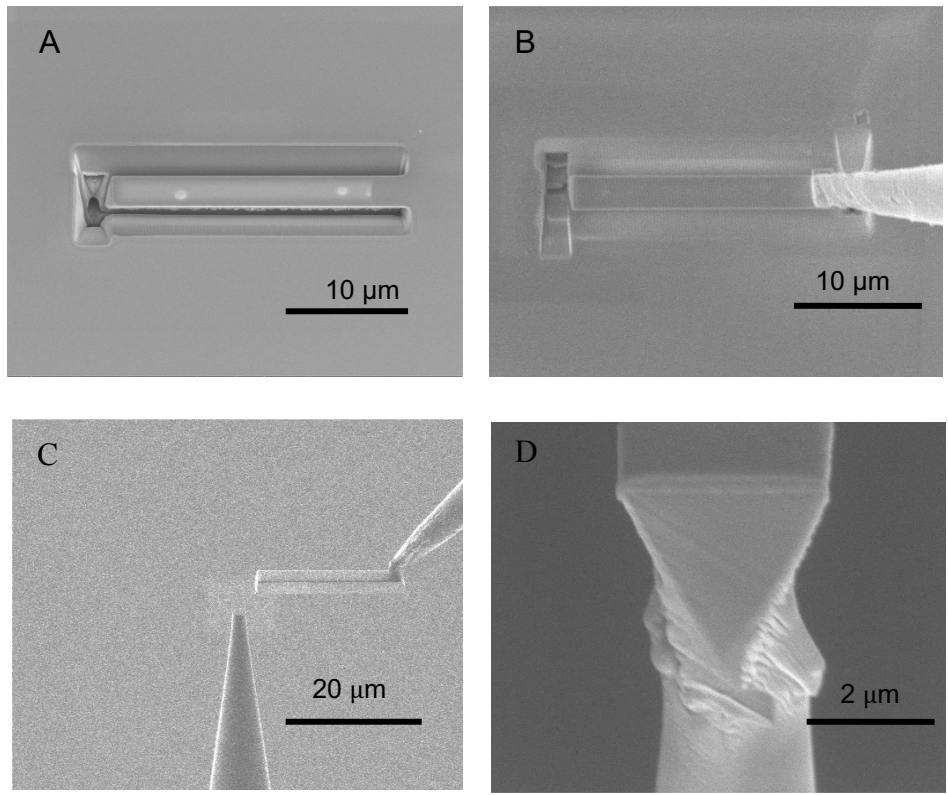

**Figure S1.** Scanning Electron Microscopy (SEM) Images of Part I of FIB fabrication of APT tips from ZVI films. (**A**) Cutting a wedge-shapes blank into the ZVI film deposited on silicon wafer [SEM conditions: 1.4nA, 5kV]. (**B**) Lifting out the blank with a micromanipulator [1.4nA, 5kV]. (**C**) Attaching the blank to a silicon micropost located on a micropost array [Ion-beam image, 28pA, 30kV] (**D**) Section of blank attached by platinum deposition to a silicon micropost on a micropost array [1.4nA, 5kV].

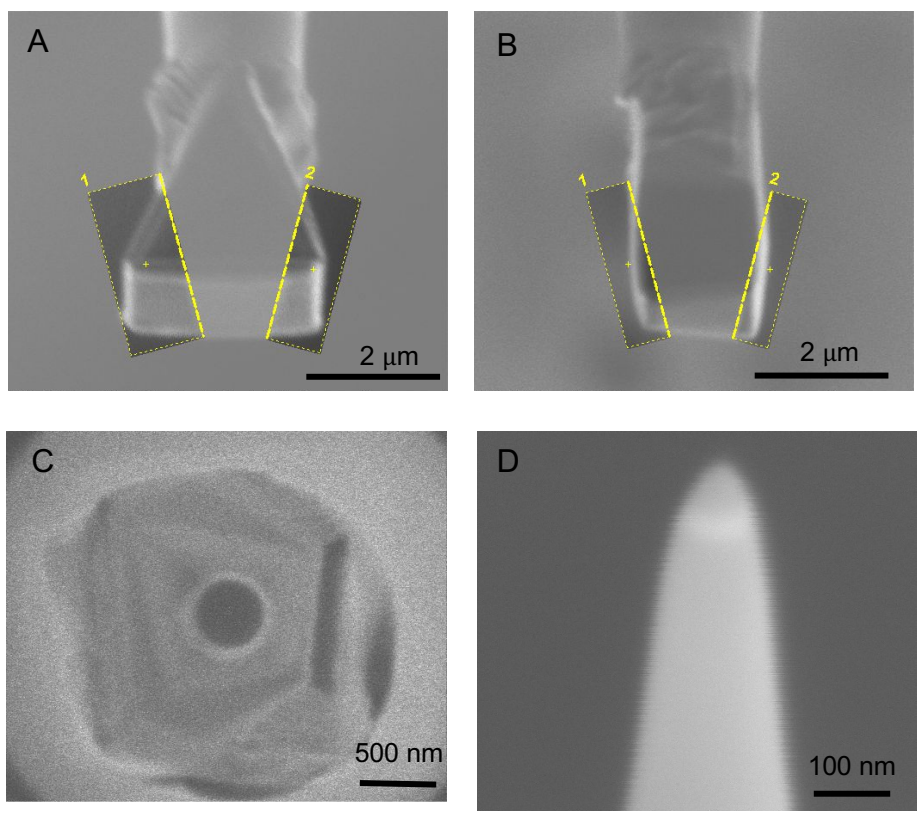

**Figure S2.** SEM Images of Part II of FIB fabrication of APT tips from ZVI films. (**A** and **B**) Triangular milling of the broad sides of the blank to obtain a pyramidal shape [0.28nA, 30kV] (**C**). Annular milling of the blank [28pA, 30kV] to shape a tip and (**D**) Low voltage circular milling to achieve the final shape of a needle-shaped ZVI tip ready for APT analysis [1.4nA, 5kV].

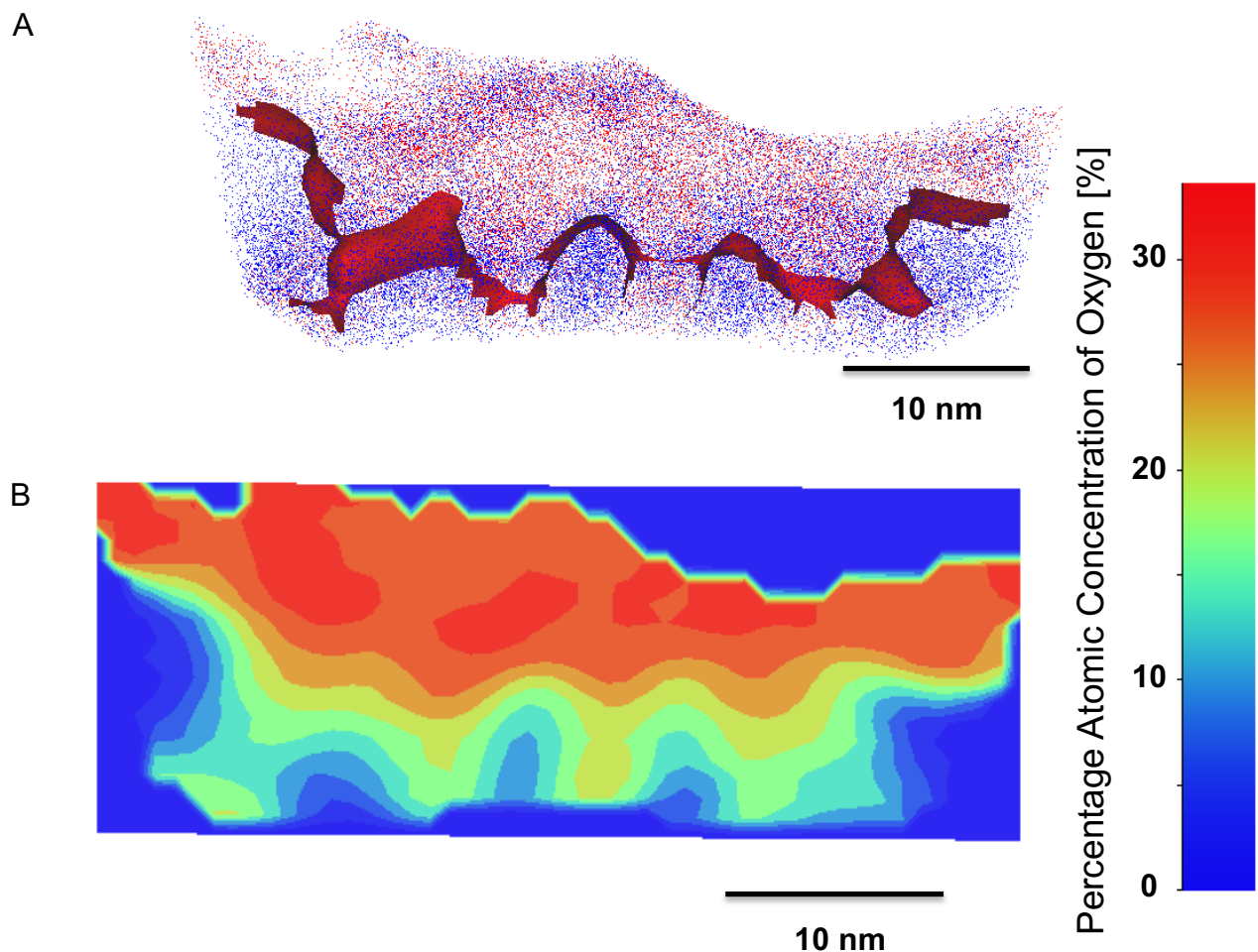

**Figure S3. (A)** 18 atomic % isoconcentration surface of $FeO^+ + Fe_2O^{2+} + O^+ + O_2^+$ (red) of a cut portion of a ZVI tip different from the one shown in Fig. 5 depicting dendritic features extending from the oxygen-rich overlayer into the bulk ZVI film at the bottom. Blue dots represent $Fe^+$, $Fe^{2+}$ and $FeH^+$ ions. Here, the ZVI tip is sliced 5 nm in the x-direction and the image is projected in the yz plane. **(B)** 2D Contour Plot of oxygen concentration for Fig. S3.A depicting the oxygen concentration gradient within the dendritic features. Color scale (see right) goes from blue to red with the red areas representing the highest concentration of $FeO^+$, $Fe_2O^{2+}$, $O^+$ and $O_2^+$.

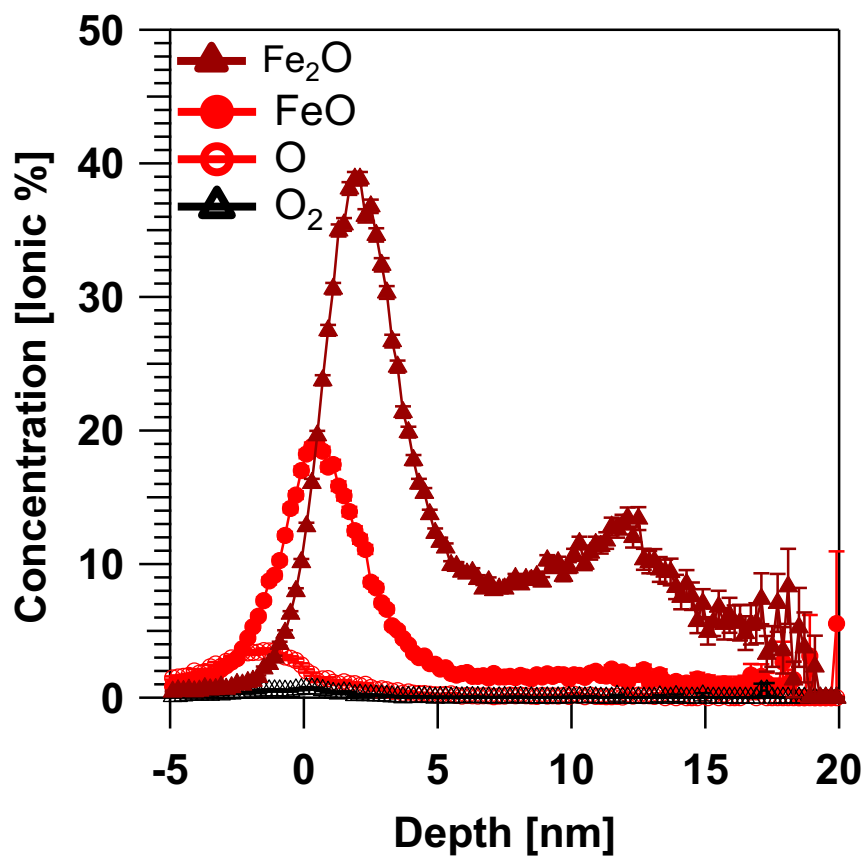

**Figure S4.** A proxigram profile of changes in ionic concentration (ionic %) of $FeO^+$, $Fe_2O^{2+}$, $O_2^+$, and $O^+$ for a typical ~ 20 nm ZVI film.